\documentclass[twocolumn,prl,aps]{revtex4}
\usepackage{floatflt}
\usepackage{psfrag}
\usepackage{amsmath}   
\usepackage{amssymb}
\usepackage{amsfonts}
\usepackage{mathrsfs}
\usepackage{graphicx}
\usepackage{xcolor}

\begin{document}
%\twocolumn [\hsize\textwidth\columnwidth\hsize\csname
%  @twocolumnfalse\endcsname
\title
{A tunable plasmonic resonator using kinetic 2D inductance and patch capacitance}

\author{V.~M.~Muravev \footnote{Corresponding E-mail: muravev@issp.ac.ru}, N.~D.~Semenov, I.~V.~Andreev, P.~A.~Gusikhin, I.~V.~Kukushkin}
\affiliation{Institute of Solid State Physics, RAS, Chernogolovka, 142432 Russia}

\date{\today}

\begin{abstract}
We have studied microwave response of a high-mobility two-dimensional electron system (2DES) contacted by two side electrodes. Using kinetic inductance of the 2DES and inter-electrode capacitance, we have constructed a subwavelength 2D plasmonic resonator. We have shown that the resonant frequency of this circuit can be controlled by 2D electron density, external magnetic field, or size of the electrodes. This opens up possibilities for using arrays of plasmonic circuits as tunable components in different frequency ranges.

\end{abstract}

\maketitle

Metamaterials are artificial materials that have extraordinary electromagnetic properties~\cite{Smith:00, Pendry:00}. The metamaterials are build from assemblies of multiple subwavelength elements. Recently, there has been proposed a new approach to the metamaterial design~\cite{Engheta:07}. Analogous to the principle used in microelectronics, it treats subwavelength elements in terms of "lumped" circuit components, such as capacitors, inductors, and resistors. This conception of metamaterial-inspired optical nanocircuitry, dubbed ''metatronics'', has been implemented successfully in experimental work conducted in terahertz and infrared frequency ranges~\cite{Averitt:06, Engheta:12, Engheta:13, Jiang:14}. 

The conductivity of 2DES at zero magnetic field can be described by the Drude formula
\begin{equation}
\sigma (\omega) = \frac{n_s e^2 \tau}{m^{\ast}} \frac{1}{1 + i \omega \tau},
\label{drude}
\end{equation}
where $n_s$ is the 2D electron density, $m^{\ast}$ is the effective mass, and $\tau$ is the transport relaxation time. Hence, in the ''lumped'' element concept, the impedance of the 2D plasma can be decomposed as~\cite{Eisenstein} 
\begin{equation}
Z_{\rm 2DES} (\omega) = R + i \omega L_{\rm K}, \qquad L_{\rm K} = \frac{m^{\ast}}{n_s e^2}.
\label{drude}
\end{equation}
Here, $L_{\rm K}$ is the kinetic inductance per square of non-magnetic origin that arises from the inertia of the electrons. In essence, the distributed kinetic inductance and capacitance of the 2DES~\cite{Aizin:12, Aizin:13} give rise to the usual two-dimensional plasmon excitation~\cite{Grimes:76, Allen:77, Theis:77}.

The objective of the presented work was to realize a subwavelength resonator circuit based on a piece of the 2D electron channel, acting as a non-magnetic inductor $L_{\rm K}$, placed in a slot between two highly conducting contacts, acting as a capacitor of capacitance $C$ (inset of Fig.~1(b)). Our experiments reveal that along with the plasmon resonance of the 2DES, such a metatronic circuit has an additional resonant response at the frequency $\omega_{\rm LC} = 1/\sqrt{L_{\rm K}C}$. Discovered resonance corresponds to the excitation of a relativistic plasmon~\cite{Muravev:15, Gusikhin:15, Muravev:20} in the circuit under study. We demonstrate that the resonant circuit can be tuned over a wide frequency range by adjusting any of the following: 2D electron density, the external perpendicular magnetic field, or the size of the contact electrodes. Due to their subwavelength confinement, the proposed 2D plasmonic structures can be used to design different 2D plasmonic devices and metamaterials~\cite{Tsui:80, Dyakonov:93, Mikhailov:98, Peralta:02, Knap:02, Knap:04, Shaner:05, Otsuji:08, Ham:12, Ham:12-2, Sirtori:15, Mu:19, Li:20, Tan:20}.

Our measurements were performed on several samples made from GaAs/AlGaAs heterostructure with $30$~nm wide quantum well situated at a depth of $h = 440$~nm below the crystal surface. The electron density was $n_s = 2 \times 10^{11}~\text{cm}^{-2}$ with the low-temperature mobility reaching approximately $2 \times 10^6~\text{cm}^2/\text{V$\cdot$s}$. A wet chemical etching produces a mesa in the shape of $1.5 \times 0.5$~mm$^2$ rectangle with two Au/Ge ohmic contacts at both sides of the mesa. The contacts cover the 2DES area of $0.5 \times 0.5$~mm$^2$ (see inset to Fig.~1). A particular geometry of the contact electrodes in each experimental sample is specified further in the discussion. The semiconductor chip had a thickness of $0.55$~mm. Microwave radiation in the frequency range $1-50$~GHz was guided to the sample through a coaxial cable or WR $62$ waveguide. To measure the absorption of microwave radiation, we utilized a   non-invasive optical technique~\cite{Ashkinadze, Muravev:16}. It is based on the extreme thermal sensitivity of the 2DES photoluminescence spectrum. The photoluminescence spectra from 2DES with and without microwave radiation were recorded by means of the spectrometer (Coderg PHO) with a liquid-nitrogen cooled charge-coupled device camera (Princeton Instruments CCD). After that, we calculated an integral of the absolute value of the differential spectrum. This value was used as a measure of microwave absorption intensity of the sample. Experiments were performed in the liquid helium with a superconducting magnet at a base temperature of $T=4.2$~K. The magnetic field ($0 - 0.5$~T) was directed perpendicular to the sample surface.

\begin{figure}[!t]
\includegraphics[width=\linewidth]{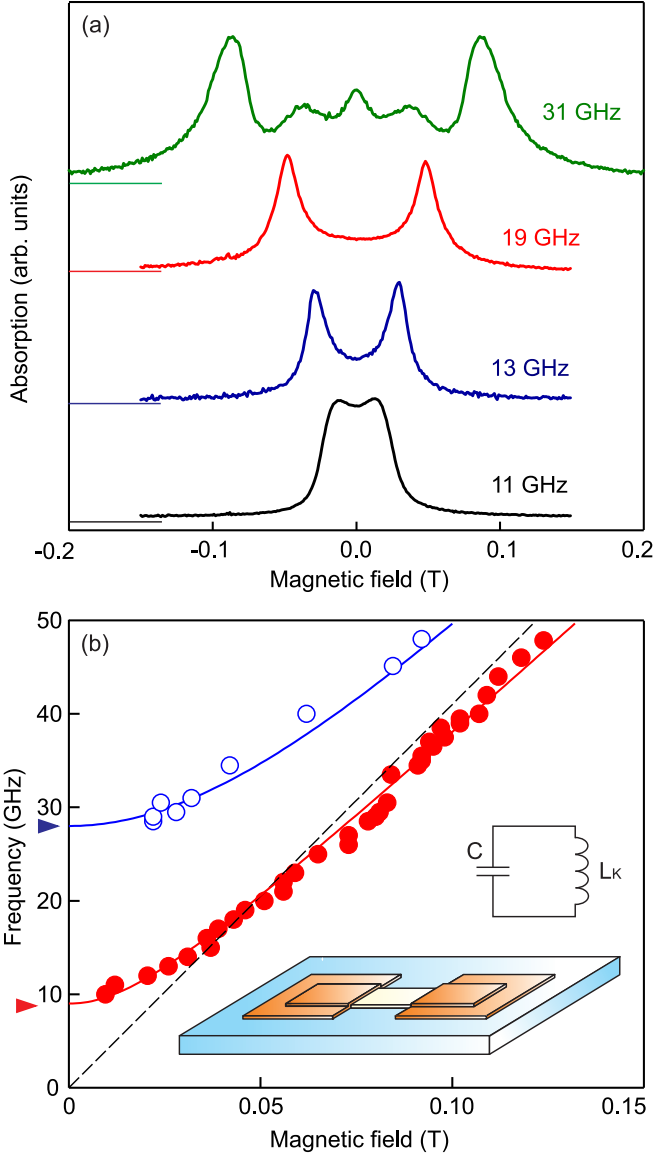}
\caption{(a) Microwave absorption as a function of the magnetic field. The curves are offset for clarity. Horizontal lines mark the zero signal level. (b) Magnetodispersion of the plasmonic $LC$ resonance (red points) and 2D plasmon (blue circles). Arrows mark theoretical prediction for plasmon frequencies at $B=0$~T. The solid lines are calculated according to Eq.~(\ref{mp_pyth}). The inset displays a schematic drawing of the device under study with its equivalent resonant circuit.}
\label{fig1}
\end{figure} 

Figure~\ref{fig1}(a) illustrates typical microwave absorption spectra measured at $11$, $13$, $19$ and $31$~GHz as a function of the externally applied magnetic field, $B$. These data were obtained for the sample with $0.85 \times 0.7$~mm$^2$ contact size. The first resonance arises after $10$~GHz. Below this frequency we do not observe any resonance response. Above $30$~GHz two resonances are observed at the same time. Figure~\ref{fig1}(b) displays the resultant magnetodispersion of the plasma modes.  We assert that the upper curve (empty points) corresponds to the excitation of the usual 2D plasmon in the $0.5 \times 0.5$~mm$^2$ 2DES square sandwiched between the contacts. The lower curve (solid red points), on the other hand, represents the magnetodispersion of a mode of the 2D plasmonic circuit. This picture is corroborated in the remainder of the manuscript. Both magnetodispersions can be accurately described by a quadratic relation resulting from hybridization between the cyclotron and plasma motions of electrons~\cite{Chaplik:72} 
\begin{equation}
\omega^2 = \sqrt{\omega_{\rm c}^2 + \omega_{\rm p}^2},
\label{mp_pyth}
\end{equation}
where $\omega_c=eB/m^{\ast}$ is the cyclotron resonance (CR) frequency and $\omega_{\rm p}$ is the plasmon frequency at zero magnetic field. The expression~(\ref{mp_pyth}) can be obtained using "lumped" element approach. Indeed, the Drude conductivity in the presence of the magnetic field reads  
$$ \sigma_{xx} (\omega) = \frac{n_s e^2 \tau}{m^{\ast}} \frac{1 + i \omega \tau}{(1 + i \omega \tau)^2 + \omega_c^2 \tau^2}$$.
By decomposing the impedance $Z_{\rm 2DES} = 1/ \sigma_{xx}$ into its active and reactive components in the limit of $\omega \tau \gg 1$, we arrive at 
$$L_{\rm K} (B) = \frac{m^{\ast}}{n_s e^2} \left( 1- \frac{\omega_c^2}{\omega^2} \right).$$
Then, taking into account that $\omega = 1/\sqrt{L_{\rm K} (B)C}$ and $\omega_p = 1/\sqrt{L_{\rm K} (B=0) C}$, the expression above reduces to the magnetodispersion relation in (\ref{mp_pyth}). In our experiment, the $LC$ circuit magnetoplasma mode intersects the CR line. This is caused by the retardation effects, which reduce the effective cyclotron frequency~\cite{Kukushkin:03, Muravev:20}.

To ascertain the origin of the observed plasma modes, we calculate the expected plasmon frequencies. 2D plasma wave at zero magnetic field possess the following dispersion law~\cite{Stern}:
\begin{equation}\label{eq_fun}
    \omega_p (q) = \sqrt{\frac{n_\text{s} e^2 q}{2 \varepsilon_0 \varepsilon^{\ast} m^{\ast}}}, \end{equation}
where $\varepsilon^{\ast} = (\varepsilon_{\rm GaAs}+1)/2$ is the effective permittivity of the surrounding medium, and $\varepsilon_0$ is the electric constant. Hence, for the plasmon wave vector $q =\pi /W$ ($W=0.5$~mm), Equation~(\ref{eq_fun}) yields the 2D plasmon frequency $f_p=28$~GHz (arrow in Fig.~1(b)), which is in excellent agreement with experimental data for the high-frequency mode denoted by empty circles in Fig.~1(b). It is important to note that Eq.~(\ref{eq_fun}) can be expressed as follows $\omega_p (q) = 1/\sqrt{L_{\rm K} C_p(q)}$, where $C_p(q)=2 \varepsilon_0 \varepsilon^{\ast}/q$ is $q$-dependent distributed self-capacitance of the 2DES~\cite{Aizin:12, Aizin:13}.

\begin{figure}[!t]
\includegraphics[width=\linewidth]{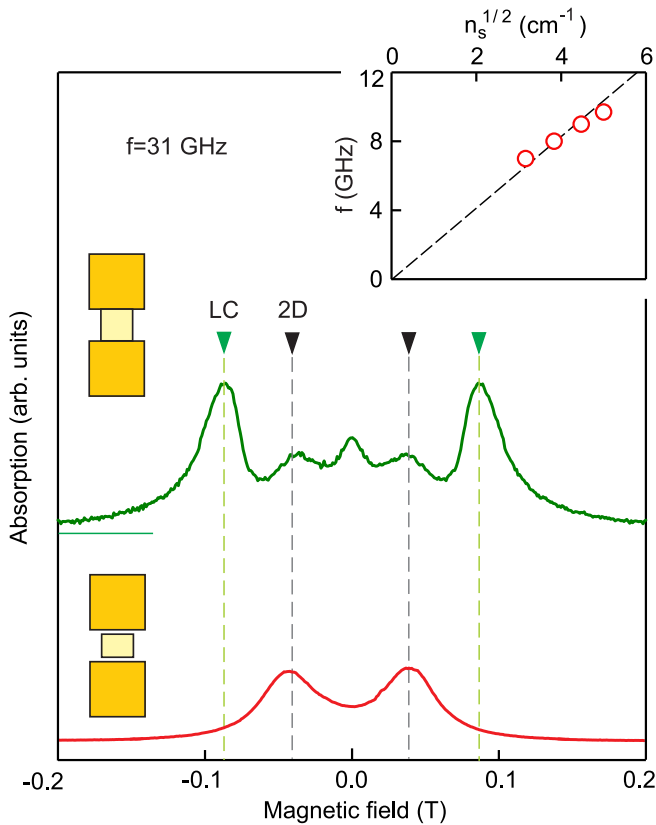}
\caption{Microwave absorption at $31$~GHz measured for the native (upper curve) and open (bottom curve) plasmonic circuits. Schematic drawings indicate the plasmonic resonator geometry corresponding to each curve. The inset displays the resonant frequency of the plasmonic $LC$ circuit as a function of the electron density. The dashed line is a guide for the 
eye.}
\label{fig2}
\end{figure} 

Now we turn our attention to the low-frequency mode of Fig.~1(b). Let's evaluate the frequency of the $LC$ plasmonic circuit realized in our experiment $\omega_{\rm LC} = 1/\sqrt{(L_{\rm K} + L) C}$, where $L$ represents the magnetic contribution to the net inductance, due to the retardation effects. This additional term plays an important role for a particular sample geometry under study. For the given sample configuration, numerical simulations yield the capacitance between the contact plates, $C = 0.115$~pF, and magnetic inductance $L= 1.7$~nH (see Supplemental Material for more details), while Eq.~(\ref{drude}) gives the kinetic inductance of the square 2DES, $L_{\rm K} = 1.2$~nH. Hence, we get the resonant frequency $f_{\rm LC}=8.7$~GHz, which is in good agreement with the experimental result of $f_{\rm exp}=9$~GHz. We note that the observed $LC$ plasma mode is equivalent to the relativistic plasmon discovered recently in the gated 2DES~\cite{Muravev:15, Gusikhin:15, Muravev:20}. In these studies, it was detected only when the gate was electrically connected to the 2DES through a wire. This condition ensures that the oscillatory plasmon circuit is closed in terms of current. In our investigation, we explore a plasmonic $LC$ circuit where the electric current flowing through the 2DES and the displacement current induced between the contact plates together form a closed loop. 

To further confirm that the observed low-frequency mode originates from oscillations in the plasmonic $LC$ resonator, we effectively opened the circuit by etching $50$~$\mu$m strips of mesa alongside each of the contacts (Fig.~\ref{fig2}). As a result of breaking the closed loop of current flow, the $LC$ plasma mode disappeared, whereas the 2D plasmon mode remained essentially unaffected. Examples of microwave absorption at $31$~GHz measured for the closed and open plasmonic circuits are depicted in Fig.~\ref{fig2} as the top and bottom curves, respectively. 

One of the most outstanding properties of the plasmonic $LC$ resonator is that its frequency can be tuned over a wide range by changing the 2D electron density. At the same time, the overall size of the resonator ($2.2 \times 0.7$~mm$^2$) is much less than the radiation wavelength ($30$~mm). Therefore, the plasmonic circuit can act as an element of active metamaterial devices. In our experiments, we tuned the 2D electron density in the sample structure depicted in Fig.~\ref{fig1} using a photodepletion technique~\cite{Kukushkin:89}. Data in the inset to Fig.~\ref{fig2} shows that when electron density is changed from $n_s=2.5 \cdot 10^{11}$~cm$^{-2}$ to $1.0 \cdot 10^{11}$~cm$^{-2}$, plasmon frequency at $B=0$~T of the plasmonic $LC$ resonator demonstrate a corresponding decrease. The dashed line in the inset is provided as a guide for the eye.

\begin{figure}[!t]
\includegraphics[width=\linewidth]{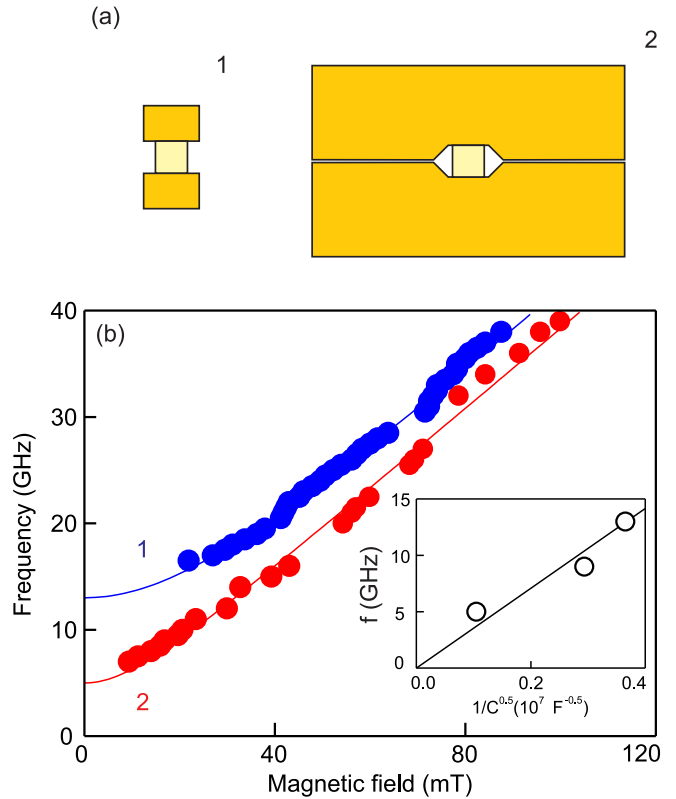}
\caption{(a) Schematic drawing of two samples with different contact areas and fixed 2DES dimensions. (b) Magnetodispersion of the $LC$ plasma mode observed for each sample. The inset displays the $LC$ plasmon frequency versus the reciprocal square root of the capacitance, which was calculated numerically. The solid curve shows a linear fit to the experimental data.}
\label{fig3}
\end{figure}

The capacitance of the $LC$ plasmonic resonator is associated with the electric field induced by opposite charges accumulated on the surface of two contacts. Therefore, the contact geometry and size affect the capacitance $C$ and, consequently, the frequency of the plasmonic resonator, $f_{\rm LC}$. To verify this experimentally, we fixed the size $0.5 \times 0.5$~mm$^2$ of the 2DES from our previous experiments and varied the contact geometry. Thus, two additional samples were tested, with contact sizes of $0.4 \times 0.7$~mm$^2$ (sample 1) and $1.1 \times 3.9$~mm$^2$ (sample 2). The schematics of these samples are included in Fig.~\ref{fig3}(a). For detailed drawings, see the Supplemental Material. Magnetodispersion of the $LC$ plasma mode observed for each sample is shown in Fig.~\ref{fig3}(b). From the data, it is evident that an increase in the contact area causes a significant decrease in the plasmon frequency. Inset to Fig.~\ref{fig3}(b) plots dependency of the plasmonic resonance frequency on $1/\sqrt{C}$. The dependency is linear in accordance with the dispersion relation $\omega_{\rm LC} = 1/\sqrt{(L_{\rm K} + L) C}$. The magnetic inductance $L$ is almost the same for all three samples.

\begin{figure}[!t]
\includegraphics[width=\linewidth]{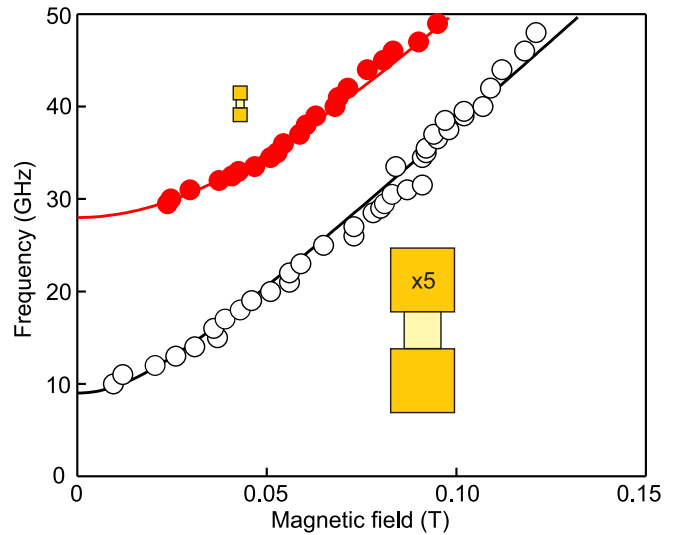}
\caption{Magnetodispersion  of  the plasmonic LC resonance for the structure with all dimensions made $5$ times smaller (red points). Empty circles display the magnetodispersion of the original sample. The insets show schematics of the devices.}
\label{fig4}
\end{figure}

The key practical advantage of the proposed device concept is that its resonant frequency is easily scalable with general structure size. To demonstrate this capability, we conducted an additional experiment with the structure geometry identical to that in Fig.~\ref{fig1}, but with all dimensions made $5$ times smaller. As a result, the device has a 2DES mesa, $0.1 \times 0.1$~mm$^2$ in area, placed between two $0.17 \times 0.14$~mm$^2$ contacts. Figure~\ref{fig4} shows the resultant magnetodispersion of the $LC$ plasma mode (filled circles), which indicates that downscaling leads to a significant increase in the resonant frequency $f_{\rm LC}$ from $9$ to $28$~GHz. For comparison, the 
calculated frequency $\omega_{\rm LC} = 1/\sqrt{L_{\rm K}  C}$ for the scaled device is $30.3$~GHz. In this case, the magnetic inductance for the given structure has a negligible effect. Therefore, the use of micrometer-sized devices should enable boosting the frequency of the plasmonic $LC$ circuit into the terahertz frequency range.

In conclusion, we have investigated resonant microwave absorption in the two-dimensional electron system that is contacted by two-side patch electrodes. We have found that such a device supports two types of plasma modes. One of them is an ordinary 2D plasmon excited in the 2DES. Unexpectedly, the other mode corresponds to resonant excitation of the plasmonic $LC$ circuit, constructed from the 2DES acting as a non-magnetic inductor and inter-electrode capacitor. We have demonstrated that discovered plasmonic resonator can be readily engineered by geometric shaping of either contacts or 2DES. Thanks to its subwavelength confinement and tunability, the plasmonic $LC$ resonator can be used as an element in active planar metamaterials. The presented proof-of-concept devices were implemented with GaAs 2DES and operated at GHz frequencies at cryogenic temperature. However, it has been shown that room temperature excitation of plasma waves is possible at THz and infrared wavelengths with both GaAs 2DESs and graphene~\cite{Knap:04, Sirtori:15, Ju:11, Basov}. Thus, the proposed device concept can be extended to these higher frequencies to enable room-temperature operation. 

\subsection{Supplementary Material}

Supplementary Material I includes details on Capacitance and Inductance modeling.
Supplementary Material II provides drawings for structures used in the experiments.

\subsection{Acknowledgments}

We thank V.A.~Volkov and A.A.~Zabolotnykh for the stimulating discussions. This work was supported by the Russian Science Foundation (Grant No.~18-72-10072).

The data that support the findings of this study are available from the corresponding author upon reasonable request.

\end{document}